\documentclass[twocolumn]{aastex61}

\usepackage{aas_macros}          % package to read AAS bib file
\usepackage{nccmath}             % package for fleqn
\usepackage{mathtools}           % package for dcases
\usepackage{cases}               % package for subnumcases 
\usepackage{tikz} 
\usepackage{blindtext} %\white{\blindtext}

\newcommand\beq{\begin{equation}}   
\newcommand\eeq{\end{equation}}  
\def\bal#1\eal{\begin{align}#1\end{align}}  % shortcuts \bal & \eal for \begin{align} and \end{align}
\newcommand\bse{\begin{subequations}}    % shortcut \beq for \begin{subequations}
\newcommand\ese{\end{subequations}}       % shortcut \eeq for \end{subequations}

\def\bml#1\eml{\begin{multline}#1\end{multline}}  % shortcuts \bml & \eml for \begin{multline} and \end{multline}

\newcommand\bga{\begin{gather}}    % shortcut \bml for \begin{gather}
\newcommand\ega{\end{gather}}      % shortcut \eml for \end{gather}

\usepackage{color}    
\newcommand\red[1]{{\color{red}#1}}

\newcommand\rdagger[1]{\red{^{\dagger}}}     

\newcommand\kelvin{\, \mathrm{K}}	
\newcommand\second{\, \mathrm{s}}	
\newcommand\persecond{\, \mathrm{s}^{-1} }	
	
\newcommand\metre{\, \mathrm{m}}

\newcommand\mbar{\, \mathrm{mbar}}	
\newcommand\gauss{\, \mathrm{G}}

\usepackage{amsbsy}        % \boldsymbol to boldfont greek symbols

\newcommand\di[1]{\mathrm{#1}}					 	%\di{} for text in eqn mode
\newcommand\trad{\tau_\di{rad} }
\newcommand\twave{\tau_\di{wave} }

\usepackage[noabbrev,capitalise]{cleveref}  %\cref{}
\newcommand\sectref[1]{Section \ref{#1}}

\newcommand\appref[1]{Appendix \ref{#1}} 
\usepackage{savesym}
\savesymbol{tablenum}
\usepackage{siunitx}
\restoresymbol{SIX}{tablenum}
\received{June 26, 2020}
\revised{June 22, 2021}
\accepted{June 30, 2021}
\submitjournal{ApJL}

\shorttitle{Observational Consequences of SWMHD on Hot Jupiters}
\shortauthors{Hindle, Bushby, \& Rogers}
\begin{document}

\title{Observational Consequences of Shallow-water Magnetohydrodynamics on Hot Jupiters}

%% you desire. In v6.1 you can include a footnote in the title.

%% The \author command is the same as before except it now takes an optional
%% arguement which is the 16 digit ORCID. The syntax is:
%% \author[xxxx-xxxx-xxxx-xxxx]{Author Name}
%%
%% This will hyperlink the author name to the author's ORCID page. Note that
%% during compilation, LaTeX will do some limited checking of the format of
%% the ID to make sure it is valid.
%%
%% Use \affiliation for affiliation information. The old \affil is now aliased
%% to \affiliation. AASTeX v6.1 will automatically index these in the header.
%% When a duplicate is found its index will be the same as its previous entry.
%%
%%
%% While authors can be grouped inside the same \author and \affiliation
%% commands it is better to have a single author for each. This allows for
%% one to exploit all the new benefits and should make book-keeping easier.
%%
%% If done correctly the peer review system will be able to
%% automatically put the author and affiliation information from the manuscript
%% and save the corresponding author the trouble of entering it by hand.
\correspondingauthor{Alex Hindle}
\email{alex.hindle@newcastle.ac.uk}

\author[0000-0001-6972-2093]{A. W. Hindle} 
\affil{School of Mathematics, Statistics and Physics, Newcastle University, Newcastle upon Tyne, NE1 7RU, UK}

\author[0000-0002-4691-6757]{P. J. Bushby}
\affil{School of Mathematics, Statistics and Physics, Newcastle University, Newcastle upon Tyne, NE1 7RU, UK}

\author[0000-0002-2306-1362]{T. M. Rogers}
\affil{School of Mathematics, Statistics and Physics, Newcastle University, Newcastle upon Tyne, NE1 7RU, UK}
\affil{Planetary Science Institute, Tucson, AZ 85721, USA }


\begin{abstract}
We use results of shallow-water magnetohydrodynamics (SWMHD) to place estimates on the minimum magnetic field strengths required to cause atmospheric wind variations (and therefore westward venturing hotspots) for a dataset of hot Jupiters (HJs), including HAT-P-7b, CoRoT-2b, Kepler-76, WASP-12b, and WASP-33b, on which westward hotspots have been observationally inferred. For HAT-P-7b and CoRoT-2b  our estimates agree with past results;  for Kepler-76b we find that the critical dipolar magnetic field strength, over which the observed wind variations can be explained by magnetism, lies between $4 \gauss$ and $19 \gauss$; for WASP-12b and WASP-33b westward hotspots can be explained by $1 \gauss$  and  $2 \gauss$ dipolar fields respectively. Additionally, to guide future observational missions, we identify $61$ further HJs that are likely to exhibit magnetically-driven atmospheric wind variations and predict these variations are highly-likely in $\sim 40$ of the hottest HJs.
\end{abstract}


%---------------------------------------------------------------------------------
\section{Introduction} \label{sec:intro}

Equatorial temperature maxima (hotspots) in the atmospheres of hot Jupiters (HJs) are generally found eastward (prograde) of the substellar point  \citep[e.g.,][]{2006Sci...314..623H,2007MNRAS.379..641C,2007Natur.447..183K,2009ApJ...690..822K}. Eastward hotspots are also archetypal in hydrodynamic simulations of synchronously-rotating HJs \citep[e.g.,][]{2002A&A...385..166S,2004JAtS...61.2928S,2005ApJ...629L..45C,2006ApJ...649.1048C} and are explained by hydrodynamic theory of wave-mean flow interactions \citep{2011ApJ...738...71S}. 

However, using three-dimensional (3D) magnetohydrodynamic (MHD) simulations, \cite{2014ApJ...794..132R} showed that HJs can exhibit 
winds that oscillate from east to west, causing east-west hotspot variations. {Using continuous {\em Kepler} data, westward venturing brightness offsets have since been identified in the atmospheres of the ultra-hot Jupiters (UHJs) HAT-P-7b \citep{2016NatAs...1E...4A} and Kepler-76b \citep{Jackson_2019}. Furthermore, thermal phase curve measurements from {\em Spitzer} have found westward hotspots on the UHJ WASP-12b \citep{2019MNRAS.489.1995B} and the cooler CoRoT-2b \citep{2018NatAs...2..220D}; and optical phase curve measurements from {\em TESS} found westward brightspot offsets on the UHJ WASP-33b \citep{2020arXiv200410767V}. Three explanations for these observations have been proposed:  cloud asymmetries confounding optical measurements \citep[][]{2013ApJ...776L..25D,2016A&A...594A..48L,2016ApJ...828...22P}; non-synchronous rotation \citep[]{2014ApJ...790...79R}; and magnetism \citep{2017NatAs...1E.131R}.} In \cite{2019ApJ...872L..27H}, we found that CoRoT-2b would need an
 implausibly large planetary magnetic field to explain its westward atmospheric winds; concluding that a non-magnetic explanation is more likely. \cite{2017NatAs...1E.131R} and \cite{2019ApJ...872L..27H} respectively used 3D MHD and shallow-water MHD (SWMHD) simulations to show that magnetism resulting from a $B_\di{dip}\gtrsim 6 \gauss$ dipolar field strength can explain westward hotspots on HAT-P-7b,  which is expected to be tidally-locked. Moreover, dayside cloud variability has recently been ruled-out as an explanation of the westward brightness offsets on HAT-P-7b \citep{2019arXiv190608127H} and, since all these testcases have near-zero eccentricities, they are expected to be  synchronously rotating. %Hence, a magnetic explanation  
 
 In this work we apply results from \cite{2021arXiv210707515H} on a dataset of HJs to calculate estimates of the minimum magnetic field strengths required to drive reversals. These conditions can be used to constrain the magnetic field strengths of UHJs.

\section{Reversal condition from shallow-water MHD} \label{sect:reverse:SWMHD}
The hottest HJs have weakly-ionised atmospheres, strong zonal winds, and are expected to host dynamo-driven deep-seated planetary magnetic fields. If a HJ's atmosphere is sufficiently ionised, winds become strongly coupled to the planet's deep-seated magnetic field, inducing a strong  equatorially-antisymmetric  toroidal field that dominates the atmosphere's magnetic field geometry \citep{2012ApJ...745..138M,2014ApJ...794..132R}. 

In hydrodynamic (and weakly-magnetic) systems, mid-to-high latitude geostrophic circulations cause a net west-to-east equatorial thermal energy transfer, yielding eastward hotspots, and net west-to-east angular momentum transport into the equator from higher latitudes, driving superrotating equatorial jets \citep{2011ApJ...738...71S}. In \cite{2021arXiv210707515H}, we showed that the presence of a strong equatorially-antisymmetric toroidal field obstructs these energy transporting circulations and results in reversed flows with westward hotspots. The threshold for such reversals can be estimated using \citep{2021arXiv210707515H}:
\bse
\begin{align}
& V_{\mathrm{A},\mathrm{crit}}  \approx  \max\left(  V_{\mathrm{A},0},  V_{\mathrm{A},f} \right),
\\& \frac{V_{\mathrm{A},0}}{c_g} = \frac{\beta/c_g}{1/R^2 + 3\beta/c_g} = \frac{(R/L_\di{eq})^{2}}{1 + 3(R/L_\di{eq})^{2}},
\\& \frac{V_{\mathrm{A},f}}{c_g} =  \alpha \left(\frac{\Delta h_\di{eq}}{H}\right)\left( \frac{\trad}{\twave}\right)^{-1} \left(\frac{2 \Omega \twave^2}{\trad} + 1 \right)^{-1} \hspace{-1em}, 
\end{align}
\label{eqn:VAcrit}
\ese
where $V_{\mathrm{A},\mathrm{crit}}$ is the reversal threshold of the toroidal field's Alfv\'en speed, with $V_{\mathrm{A},0}$ and $V_{\mathrm{A},f}$ respectively denoting the thresholds in the zero-forcing-amplitude limit and for a moderate-to-strong pseudo-thermal forcing. Here $R$ is the planetary radius, $c_g$ is the shallow-water gravity wave speed, $\beta= 2 \Omega / R$ is the latitudinal variation of the Coriolis parameter at the equator (for the planetary rotation frequency $\Omega$), $L_\di{eq}\equiv(c_g/\beta)^{1/2}$ is the equatorial Rossby deformation radius, $\alpha = 2 \pi R / L_\di{eq}$ is a longitude-latitude lengthscale ratio, $\twave \equiv L_\di{eq}/c_g$ is the system's characteristic wave time scale \citep[as in][]{2011ApJ...738...71S}, and $\Delta h_\di{eq}/H$ determines the magnitude of the shallow-water system's pseudo-thermal forcing profile, for a Newtonian cooling treatment with a  radiative timescale, $\trad$.

\begin{figure*}
 \centering %[scale=0.985]
\includegraphics[scale=0.87]{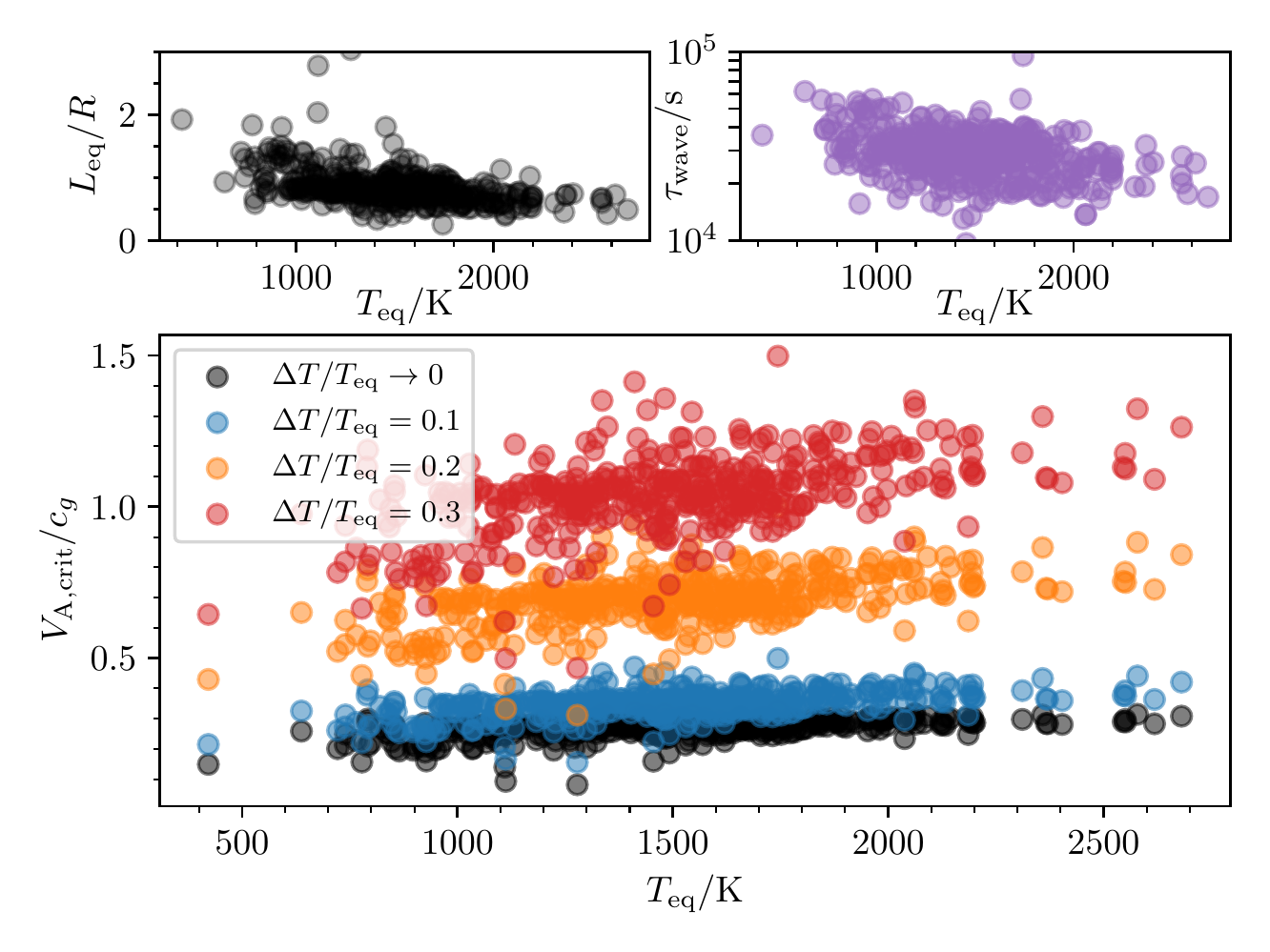} %[width=0.75\linewidth]
 \caption{ $L_\di{eq}/R$ (top left), $\twave$ (top right), and $V_{\mathrm{A},\mathrm{crit}}/c_g$ (bottom panel) vs. $T_\mathrm{eq}$, using the \href{www.exoplanet.eu}{exoplanet.eu} dataset, where  $V_{\mathrm{A},\mathrm{crit}}/c_g$ is calculated for  $\Delta T /T_\di{eq} = \{0,0.1,0.2,0.3\}$. 
 } 
 \label{fig:twave:VA}  
\end{figure*}

\section{Method for Placing Magnetic Reversal Criteria on hot Jupiters} 
%Teq from data
\cref{eqn:VAcrit} shows that the parameters $R$, $c_g$, $\Omega$, $\trad$, and $\Delta h_\di{eq}/H$ can be used to estimate the minimum magnetic field strengths required for reversals. We apply this simple relation to a dataset of HJs taken from \href{www.exoplanet.eu}{exoplanet.eu}\footnote{Accessed May 30, 2021. HJs without data entries for $R$, $M$, $t_\di{orbit}$, $a$, $e$, $R_*$, or $T_*$ are removed.}, using  planets with  $0.1 \,M_\di{J}<M<10 \,M_\di{J}$ and $a < 0.1 \, \di{AU}$,  where $M$ and $M_\di{J}$ denote the planetary mass and Jupiter's mass respectively, and $a$ is the semimajor axis. The criteria are calculated using the equilibrium temperature \citep[assuming zero albedos; e.g.,][]{Laughlin_2011}:
\beq
T_\di{eq} = \left( \frac{R_*}{2a} \right)^{1/2} \frac{T_*}{(1- e^2)^{1/8}}, \label{eqn:teq}
\eeq
for stellar radius, $R_*$, orbital eccentricity, $e$, and stellar effective temperature, $T_*$.

 The validity of the shallow-water approximation can be assessed by comparing $L_\di{eq}$ to the pressure scale height, $H \sim \mathcal{R} T_\mathrm{eq} R^2/G M $, where $G$ is Newton's gravitational constant and $\mathcal{R}$, the specific gas constant, is calculated using the solar system abundances in \cite{10.1007/978-3-642-10352-0_8}.
For the sampled HJs, $\di{mean}(H/L_\di{eq}) = \num{7.5e-3}$, so shallow-water theory is generally expected to capture their leading order atmospheric dynamics well. The shallow-water gravity wave speed is calculated by equating thermal and geopotential energies, yielding $c_g \equiv \sqrt{gH} \sim (\mathcal{R} T_\mathrm{eq})^{1/2}$. 
 Doing so implies $\Delta h/H \sim \Delta T/T_\di{eq}$, where $\Delta h$ are deviations in shallow-water layer thickness from the reference $H$ and $\Delta T\equiv T_\di{day}-T_\di{eq}$ for the dayside temperature, $T_\di{day}$. Though not exactly equal, $\trad \sim \twave$ in the upper atmospheres of hot Jupiters  \citep{2008ApJ...678.1419F,2014ApJ...794..132R,2017NatAs...1E.131R}. Taking $\trad=\twave$ is also convenient for this analysis as, when $\trad \lesssim \twave$, $\Delta h \sim \Delta h_\di{eq} $ \citep{2013ApJ...776..134P,2021arXiv210707515H}, so $\Delta h_\di{eq}/H \sim  \Delta T/T_\di{eq}$. While this treatment is a dynamic simplification, in \cite{2021arXiv210707515H} we found that it predicts reversal criteria consistent with the 3D MHD simulations of  \cite{2014ApJ...794..132R} and \cite{2017NatAs...1E.131R}.

An interesting feature of HJs is that the dynamical parameters $c_g$, $\Omega$ and $R$ of a HJ are all related to its host star proximity and the mass/radius/luminosity of its host star (i.e., they are all related to $T_\di{eq}$). The consequence of this interdependence is that, for the hottest HJs, $L_\di{eq}/R$ and $\twave$ approximately converge to $L_\di{eq}/R\approx0.7$ and $\twave \approx 2 \times 10^4 \second$ (see  \cref{fig:twave:VA}; top panels).  In \cref{fig:twave:VA} (bottom panel) we use \cref{eqn:VAcrit} to plot  $V_{\mathrm{A},\mathrm{crit}}/c_g$ for $\Delta T/T_\di{eq} = 0, 0.1, 0.2, 0.3$. Taking $\Delta T \approx (T_\di{day}-T_\di{night})/2$, $\Delta T/T_\di{eq} = 0.1, 0.2, 0.3$  cover the expected range of relative dayside-nightside variations  \citep[e.g.,][]{2017ApJ...835..198K}; whereas $\Delta T/T_\di{eq} = 0$ shows the zero-amplitude limit. $V_{\mathrm{A},\mathrm{crit}}/c_g$ varies linearly with $\Delta T/T_\di{eq}$ above $\Delta T/T_\di{eq} = 0.1$, but approaches the zero-amplitude limit for $\Delta T/T_\di{eq} \lesssim 0.1$. A remarkable feature of the HJ dataset is that, due to the aforementioned interdependences, the ratio $V_{\mathrm{A},\mathrm{crit}}/c_g$ also converges in the large $T_\di{eq}$ limit for a given $\Delta T/T_\di{eq}$. 

\cref{eqn:VAcrit}, the Alfv\'en speed definition, and the ideal gas law yield
\beq 
B_{\phi,\mathrm{crit}} = \left(\frac{\mu_0 P}{\mathcal{R} T}\right)^{1/2} V_{\mathrm{A},\mathrm{crit}} \sim \frac{V_{\mathrm{A},\mathrm{crit}}}{c_g} \sqrt{\mu_0 P}\   , \label{eqn:Bphi:calc}
\eeq
where $B_{\phi,\mathrm{crit}}$ is the critical threshold of the toroidal field magnitude $B_\phi$, $\mu_0$ is the permeability of free space, and $T$ and $P$ are the temperature and pressure at which the reversal occurs. 

If the electric currents that generate the planet's assumed deep-seated dipolar field are located far below  the atmosphere, \cite{2012ApJ...745..138M} showed that $B_\phi$ can be related to the dipolar field strength, $B_\di{dip}$, by the scaling law
\beq
B_\phi \sim R_m  B_\di{dip}, \label{eqn:Bphi:Bdip} 
\eeq
where  $R_m = {U_\phi H}/{\eta}$ is the magnetic Reynolds number for a given magnetic diffusivity, $\eta$, zonal wind speed, $U_\phi$, and  pressure scale height, $H$. $R_m$ estimates the relative importance of the atmospheric toroidal field's induction and diffusion; while $U_\phi/c_g$ scales linearly with $\Delta h/H \sim \Delta T/T_\di{eq}$ in geostrophically or drag dominated flows  \citep{2013ApJ...776..134P}. Taking a geostrophically-dominated flow yields $f U_\phi \sim ( \Delta T/T_\di{eq}) c_g^2/L_\di{eq}$, so $U_\phi/c_g \sim ( \Delta T/T_\di{eq}) L_D/L_\di{eq}$, with $L_D=c_g/f$. We fix the constant of proportionality in this scaling by setting $U_\phi \sim 1.5 \times 10^2 \metre \persecond$ for the conditions corresponding to the simulations of \cite{2017NatAs...1E.131R}. We calculate $\eta$ following the method of \cite{Rauscher_2013} and \cite{2014ApJ...794..132R}, taking
\beq
\eta = 230 \times 10^{-4} \,\frac{\sqrt{T}}{\chi_e} \, \di{m}^2\,\di{s}^{-1},
\eeq
where $\chi_e$ is the ionisation fraction, which is calculated using a form of the Saha equation that takes into account all elements from hydrogen to nickel. It is given by
\beq
 \chi_e = \sum_{i=1}^{28} \left( \frac{n_i}{n} \right) \chi_{e,i}\,. \label{eqn:chi:tot}
 \eeq
 In this sum the number density for each element, $n_i$, and the  ionisation fraction of each element, $\chi_{e,i}$, are calculated using
 \begin{align}
& n_i = n \left(\frac{a_i}{a_H}\right)= \frac{\rho}{\mu_m}\left(\frac{a_i}{a_H}\right),\\
 \frac{\chi_{e,i} ^2}{1-\chi_{e,i} ^2}= & n_i^{-1} \left(\frac{2 \pi m_e}{h^2}\right)^{3/2} (kT)^{3/2} \exp \left( - \frac{\epsilon_i}{kT} \right), \label{eqn:ionisation:fraction}
 \end{align}
for density $\rho$, total number density $n$, molecular mass $\mu_m$, relative elemental abundance (normalised to the hydrogen abundance) $a_i/a_H$, the electron mass $m_e$,   Plank's constant $h$, the Boltzmann constant $k$, and the elemental ionisation potential $\epsilon_i$. {To calculate $\eta$, we use the solar system abundances in \cite{10.1007/978-3-642-10352-0_8} and take $T=T_\di{eq}+\Delta T / \sqrt{2}$, the root-mean-squared temperature for a sinusoidal longitudinal temperature profile. }

\begin{figure*}
 \centering
  \includegraphics[scale=0.85]{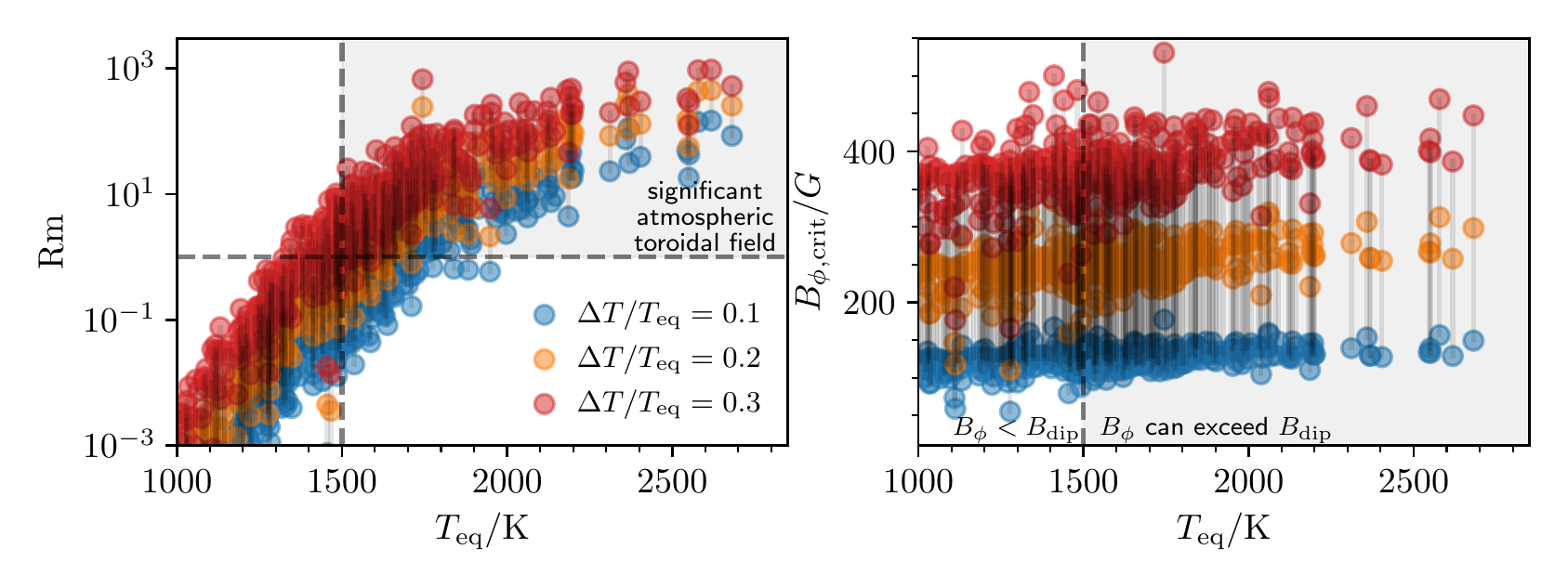}
 \caption{$R_m$ (left) and $B_{\phi,\di{crit}}$ (right) vs. $T_\di{eq}$, for the \href{www.exoplanet.eu}{exoplanet.eu} dataset. The estimates are calculated at $P=10\mbar$ with $T=T_\di{eq} + \Delta T$, where $\Delta T/T_\di{eq} = 0.1,0.2,0.3$ (blue, orange, red). For each HJ, these are connected by a translucent line. The dashed reference lines $T_\di{eq} = 1500 \kelvin$ and $R_m=1$ (lefthand panel only) are also overplotted. } \label{fig:Bphi:Rm}  
\end{figure*}

\section{Magnetic field constraints} \label{sect:Bfield}
\subsection{Estimates of $R_m$ and $B_{\phi,\di{crit}}$}

Estimates of $R_m$ and $B_{\phi,\di{crit}}$  are calculated at depths corresponding to $P=10\mbar$, at which \cite{2014ApJ...794..132R}  found magnetically-driven wind variations.  In \cref{fig:Bphi:Rm} we plot $R_m$  (lefthand panel) and  $B_{\phi,\di{crit}}$ (righthand panel) vs. $T_\di{eq}$, for HJs in the dataset (with  $T_\di{eq}>1000\kelvin$), taking $\Delta T/T_\di{eq} = 0.1, 0.2, 0.3$. 

Induction of the atmospheric toroidal field is expected to become significant when $R_m$ exceeds unity.  At $P = 10\mbar$, $R_m$ exceeds unity for $T\gtrsim1500 \kelvin$, depending on $\Delta T/T_\di{eq} $. However, due to the highly temperature dependent nature of \cref{eqn:ionisation:fraction}, 
 $R_m$ varies significantly when one compares $\Delta T/T_\di{eq} = 0.1, 0.3$ for a given HJ.  
    
 As we see in \sectref{subsect:Bdip}, $B_\phi$ is only likely to exceed $B_{\phi,\di{crit}}$ if the HJ in question is hot enough to maintain a significant atmospheric toroidal field ($R_m\gg1$). We therefore concentrate our discussion on these hotter HJs; however, we place hypothetical estimates on $B_{\phi,\di{crit}}$  for all planets in the dataset with $T_\di{eq} > 1000 \kelvin$ (\cref{fig:Bphi:Rm}, righthand panel).  Since, for a given $\Delta T/T_\di{eq}$,  $V_{\mathrm{A},\mathrm{crit}}/c_g$ is virtually independent of $T_\di{eq}$ in the hottest HJs, so is $B_{\phi,\di{crit}}$, with $100 \gauss \lesssim B_{\phi,\di{crit}} \lesssim 450 \gauss$ for $0.1<\Delta T/T_\di{eq}<0.3$; whereas larger  $L_\di{eq}/R$ values can cause $B_{\phi,\di{crit}}$  to decrease in the cooler HJs  (compare with \cref{fig:twave:VA}). We comment that $B_{\phi,\di{crit}}$ is generally least severe in the uppermost regions of the atmosphere, where the atmosphere is least dense, explaining why \cite{2014ApJ...794..132R} found the east-west wind variations at these depths. 
 
In \cite{2021arXiv210707515H}, we highlighted that magnetically-driven wind variations can be viewed as a saturation mechanism for the atmospheric toroidal field, with the reversal mechanism preventing $B_\phi$ from greatly exceeding $B_{\phi,\di{crit}}$. This suggests that $B_\phi$ should peak in the deepest regions satisfying $B_\phi \sim B_{\phi,\di{crit}}$, where $B_{\phi,\di{crit}}$ can be large, then decrease towards the surface, where $B_{\phi,\di{crit}}$ is smaller. This is consistent with \cite{2014ApJ...794..132R}, who found  $B_\phi$ peaks in the mid-atmosphere (and declined to $300 \gauss \lesssim B_{\phi} \lesssim 450 \gauss$ at $P=10\mbar$ in their M7b simulations).
 
\subsection{Dipolar magnetic field strengths} \label{subsect:Bdip}

\begin{figure*}
 \centering
  \includegraphics[scale=0.9]{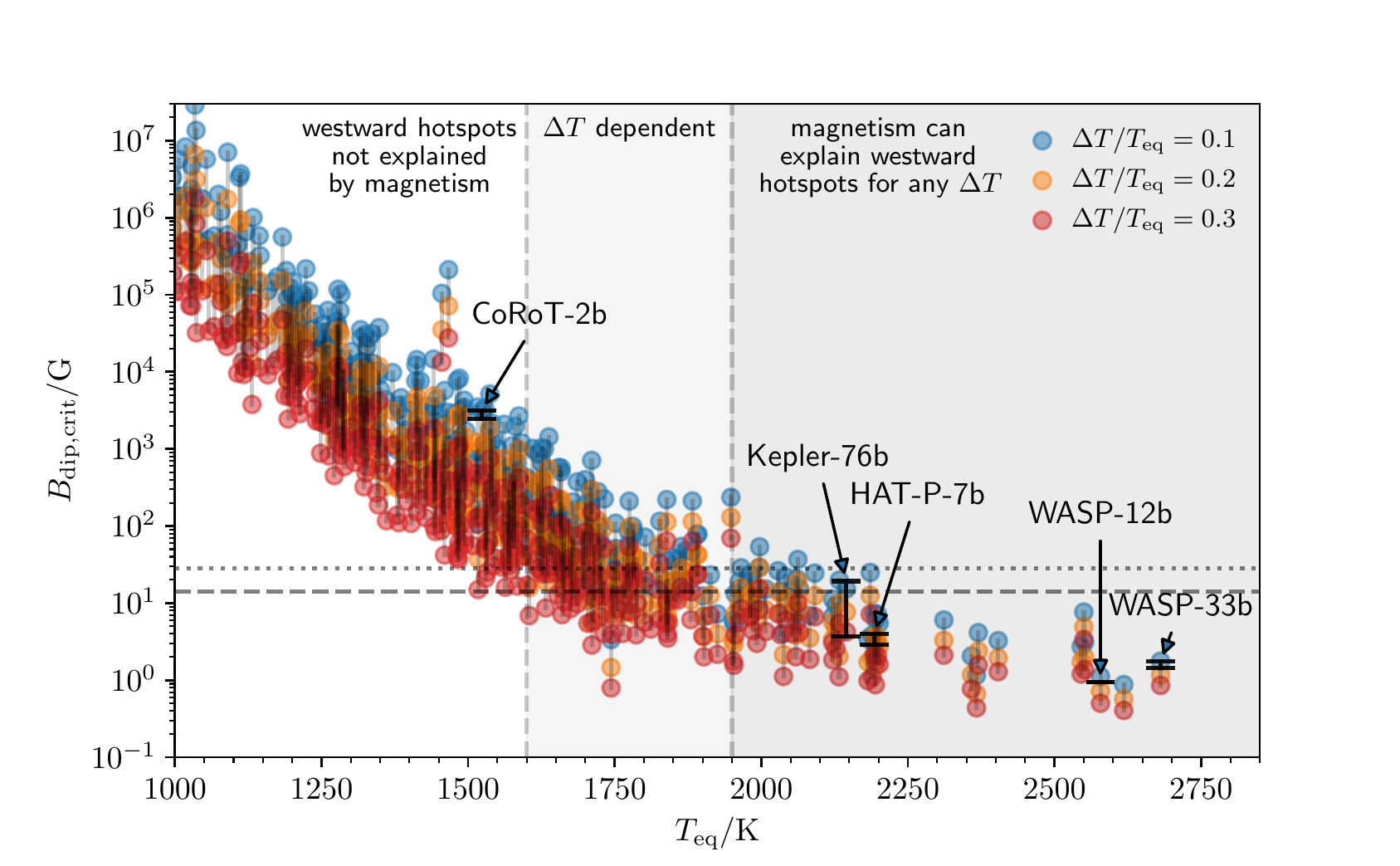} 
 \caption{Critical dipole magnetic field strengths, $B_{\di{dip},\di{crit}}$, at $P=10 \mbar$. We plot $B_{\di{dip},\di{crit}}$ using $T=T_\di{eq}+\Delta T$, with $\Delta T/T_\di{eq} = 0.1,0.2,0.3$  (blue, orange, red). For a given HJ, these are connected by translucent lines. We include error bars and labels for the planets discussed in this letter (see \cref{tab:Bphi:Bdip:estimates}) along with  reference lines at $14 \gauss$ (dashed; Jupiter's polar surface magnetic field strength) and $28 \gauss$ (dotted; twice this).
} \label{fig:Bdip}  
\end{figure*}

In \cref{fig:Bdip}  we use  \cref{eqn:Bphi:Bdip} to plot $T_\di{eq}$ vs. $B_{\di{dip},\di{crit}}$, the critical dipolar field (at $P=10\mbar$) for $\Delta T/T_\di{eq} = 0.1, 0.2, 0.3$. Since the translation of planetary dynamo theory into the HJ parameter regime is not well-understood, we include a physically motivated reference line at $B_{\di{dip},\di{crit}}=14 \gauss$ (the magnitude of Jupiter's magnetic field at its polar surface) and a second reference line at $28\gauss$ (twice this). Due to the  highly temperature dependent nature of $R_m$, these estimates of $B_{\di{dip},\di{crit}}$ carry a high degree of uncertainty (e.g., compare $B_{\di{dip},\di{crit}}$ of a given HJ for the different $\Delta T/T_\di{eq}$ choices). Therefore, for useful estimates of $B_{\di{dip},\di{crit}}$,  accurate temperature estimates/measurements (at the depth being probed) are required.

Generally, $T_\di{day}$ is not directly calculable from standard planetary/stellar parameters, so measured values should be used where possible. For the five HJs with westward hotspot observations, we use dayside temperatures based on phase curve measurements to estimate 
$B_{\phi,\di{crit}}$ and  $B_{\di{dip},\di{crit}}$. We present these estimates in \cref{tab:Bphi:Bdip:estimates} and add labelled error bars to \cref{fig:Bdip}. The UHJs are found to have low-to-moderate $B_{\di{dip},\di{crit}}$ requirements. For HAT-P-7b we estimate $3\gauss <B_{\di{dip},\di{crit}}< 4\gauss$ at $P=10\mbar$\footnote{Since $B_\di{dip}$ scales like $r^{-3}$, these estimates bracket the $B_{\di{dip},\di{crit},\di{base}} \sim 6\gauss$ prediction of  \cite{2017NatAs...1E.131R}, made for magnitudes at the atmospheric base.}, recovering the previously-known result that westward hotspots on HAT-P-7b can be well-explained by magnetism \citep{2017NatAs...1E.131R,2019ApJ...872L..27H}. On the UHJs WASP-12b and WASP-33b dipole fields respectively exceeding $1 \gauss$ and $2 \gauss$ at $P=10\mbar$ would explain westward hotspots. Likewise, at $P=10\mbar$, a dipole field exceeding $B_{\di{dip},\di{crit}}$ for  $4\gauss <B_{\di{dip},\di{crit}}< 19\gauss$ is required to explain westward hotspots on Kepler-76b. Given the comparison with Jupiter and that \cite{2019NatAs...3.1128C} predicted surface magnetic fields on HJs could range from $20 \gauss$ to $120 \gauss$, these estimates support the idea that wind reversals on these UHJs have a magnetic origin. If non-magnetic explanations can be ruled out, such estimates of $B_{\di{dip},\di{crit}}$ can be used as lower bounds for $B_{\di{dip}}$ on UHJs. In contrast, unless CoRoT-2b hosts an unfeasibly large $\gtrsim 3 \, \mathrm{kG}$ dipolar field, its westward hotspots are not explained by magnetism \cite[recovering the result of][]{2019ApJ...872L..27H}. To check our method's fidelity, we also compare predictions to the simulations in \cite{2014ApJ...794..132R}, finding good agreement (for both $B_{\di{dip},\di{crit}}$ and $B_{\di{\phi},\di{crit}}$).

 Using the range $\Delta T/T_\di{eq} = (0.1, 0.3)$ to estimate $B_{\di{\phi},\di{crit}}$ generally has uncertainties between one-half and one order of magnitude. However, \cref{fig:Bdip} shows that HJs divide into three clear categories: (i) those likely to have magnetically-driven atmospheric wind variations for any choice of $\Delta T/T_\di{eq}$ ($T_\di{eq}\gtrsim 1950 \kelvin$); (ii) those unlikely to have sufficiently strong toroidal fields to explain atmospheric wind variations, for any choice of $\Delta T/T_\di{eq}$ ($T_\di{eq} \ll 1600 \kelvin$); and (iii) marginal cases that depend on the magnitude of day-night temperature differences ($1600 \kelvin \lesssim T_\di{eq}\lesssim 1950 \kelvin$). 
 
Using the conditions $B_{\di{dip},\di{crit}} < 28 \gauss$, $P=10 \mbar$, and $\Delta T/T_\di{eq} = 0.1$, we identify 61 further HJs that are likely to exhibit magnetically-driven wind variations. We present these in \cref{tab:reversal:HJs} (\appref{app:table}), which is ordered by ascending $B_{\di{dip},\di{crit}}$ (i.e., from most-likely to least-likely to exhibit reversals), to help guide future observational missions. Of these 61 reversal candidates, 37 HJs have weaker reversal requirements than Kepler-76b. Hence, using these fairly conservative criteria, we predict that magnetic wind variations could be present in
$\sim60$ and argue that they are highly-likely in $\sim 40$ of the hottest HJs.\footnote{Using the more flexible criteria $B_{\di{dip},\di{crit}} < 28 \gauss$ at $P=10 \mbar$, with $\Delta T/T_\di{eq} = 0.2$, we find a total of 94 candidates. }

For HJs with intermediate temperatures ($1600 \kelvin \lesssim T_\di{eq}\lesssim 1950 \kelvin$), the magnitude of $\Delta T/T_\di{eq}$ (and our simplifying assumptions) plays a significant role in determining whether magnetic wind variations are plausible, so specific dayside temperature measurements should be used for estimates. These intermediate temperatures HJs offer excellent opportunities to fine-tune magnetohydrodynamic theory, via cross-comparisons between observations and bespoke models.

 \begin{deluxetable}{llll}
\tablecaption{Estimates of $B_{\phi,\di{crit}}$ and  $B_{\di{dip},\di{crit}}$ at $P=10\mbar$, using the tabulated $T_\di{day}$, for  HAT-P-7b, CoRoT-2b, Kepler-76b, WASP-12b, and WASP-33b. \label{tab:Bphi:Bdip:estimates}}
\tablecolumns{4}
\tablehead{
\colhead{Planet} & 
\colhead{$T_\di{day}/\kelvin$} & 
\colhead{$B_{\phi,\di{crit}}/\gauss$} & 
\colhead{$B_{\di{dip},\di{crit}}/\gauss$} 
}
\startdata
HAT-P-7b & $(2610,2724)^1$  & $(255,324)$ & $(3,4)$   \\      
CoRoT-2b & $(1695,1709)^2$  & $(145,177)$ & $(2500,3100)$\\   
Kepler-76b &  $(2300,2850)^3$ & $(107,466)$ & $(4,19)$ \\     
WASP-12b & $(2928)^4$   & $(212)$ & $(0.9)$ \\    
WASP-33b & $(2954,3074)^5$   & $(152,218)$ & $(1.4,1.8)$ \\    
\enddata
\tablenotetext{}{$^1$\cite{2016ApJ...823..122W}; $^2$\cite{2018NatAs...2..220D}; $^3$\cite{Jackson_2019}; $^4$\cite{2012ApJ...747...82C}; $^5$\cite{2020arXiv200410767V}.}
\end{deluxetable}

\section{Discussion}
 % Comparing reversal criteria 
We have applied the theory developed in \cite{2021arXiv210707515H} to a dataset of HJs to estimate the critical magnetic field strengths $B_{\di{dip},\di{crit}}$ and $B_{\phi,\di{crit}}$ (at $P=10\mbar$), beyond which strong toroidal fields cause  westward  hotspots. The new criterion differs both mathematically and in physical interpretation from the criterion of \cite{2014ApJ...794..132R} and \cite{2017NatAs...1E.131R}, which identifies when Lorentz forces from the deep-seated dipolar field become strong enough to significantly reduce zonal winds, but doesn't theoretically explain wind variations. However, the estimates made in this work match well with typical magnetic fields in the 3D simulations of \cite{2014ApJ...794..132R} and \cite{2017NatAs...1E.131R}, which exhibit wind variations, and also match values resulting from their criterion in these regions of parameter space. This is because, while describing different magnetic effects, both criteria  predict the critical magnetic field strengths at which magnetism becomes dynamically-important  in HJ atmospheres. Applying the new criterion to the HJ dataset, we found that the brightspot variations on Kepler-76b can be explained by plausible planetary dipole strengths ($B_\di{dip} \gtrsim 4 \gauss$ using $T_\di{day} = 2850$; $B_\di{dip} \gtrsim 19 \gauss$ using $T_\di{day} = 2300$), and that westward hotspots can be explained for $B_\di{dip} \gtrsim 1 \gauss$ on WASP-12b and $B_\di{dip} \gtrsim 2 \gauss$  on WASP-33b. The estimates of $B_{\phi,\di{crit}}$ and $B_{\di{dip},\di{crit}}$ for HAT-P-7b and CoRoT-2b are consistent with the estimates of \cite{2017NatAs...1E.131R} and \cite{2019ApJ...872L..27H}. We then used an observationally motivated set of criteria ($B_{\di{dip},\di{crit}} < 28 \gauss$, $\Delta T/T_\di{eq} = 0.1$, and $P=10 \mbar$) to tabulate {65} HJs that are likely to exhibit magnetically-driven wind variations (see \cref{{tab:reversal:HJs}}, \appref{app:table}) and predict such effects are highly-likely in $\sim 40$ of the hottest HJs.

%Motivations and limitations
With exoplanet meteorology becoming increasingly developed, the results of this study suggests that further observations of hotspot variations in UHJs should be expected. A combination of archival data and future dedicated observational missions from {Kepler}, {Spitzer}, {Hubble}, TESS, CHEOPS, and JWST can be used to identify magnetically-driven wind variations and other interesting features at different atmospheric depths. In particular, long time-span studies observing multiple transits of UHJs are likely to be essential in understanding hotspot/brightspot oscillations. Of the studies that have measured westward hotspot/brightspot offets, only the long time-span studies of \cite{2016NatAs...1E...4A} (HAT-P-7b; 4 years) and \cite{Jackson_2019} (Kepler-76b; 1000 days) identify hotspot/brightspot oscillations. In both cases, such oscillations are observed on timescales of $\sim$$10$-$100 \mbox{ Earth days}$, which \cite{2017NatAs...1E.131R} noted is consistent with timescales of wind variability in 3D MHD simulations (and the deep-seated magnetic field's Alfv\'en timescale). Such timescales are of-order or longer than the total time-spans of the other UHJ studies with westward hotspot/brightspot measurements \citep{2019MNRAS.489.1995B,2020arXiv200410767V}, so it is impossible to tell whether these measurements are part of an oscillatory evolution. 

If non-magnetic explanations can be ruled-out for past and future identifications of  westward hotspot offsets on UHJs, the coolest planets with wind variations can indicate typical $B_{\di{dip}}$ magnitudes on HJs. This has the potential to drive new understanding of the atmospheric dynamics of UHJs and provide important observational constraints for dynamo models of HJs. Parallel to this, future theoretical work can refine estimates of $B_{\di{dip},\di{crit}}$. In many cases combining observational measurements with bespoke 3D MHD simulations offer the best prospect for providing accurate constraints on the magnetic field strengths of UHJs, yet the simple concepts and results of this work can provide useful starting points for such studies and can highlight trends from an ensemble viewpoint. The largest limiting factor in our estimates of $B_{\di{dip},\di{crit}}$ is the highly temperature dependent nature of $R_m$. Furthermore, the magnetic scaling law does not account for longitudinal asymmetries in the magnetic diffusivity or the dipolar field strength within the atmospheric region. In future work we shall investigate how these inhomogeneities effect the atmospheric dynamics more closely, using a 3D model containing variable magnetic diffusivity, consistent poloidal-toroidal field coupling, stratification, and thermodynamics. To date, MHD models of HJs have strictly considered dipolar magnetic field geometries for the planetary magnetic field. Dynamo simulations would offer insight into the nature of magnetic fields in the deep interiors of HJs, which, at present, is not well-understood.

%---------------------------------------------------------------------------------
\acknowledgments
We acknowledge support from STFC for A.~W.~Hindle's studentship (ST/N504191/1) and the Leverhulme grant RPG-2017-035. We thank Andrew Cumming and Natalia G\'omez-P\'erez for useful conversations leading to the development of this manuscript.

\bibliography{SWMHD_bib.bib}

\appendix
\section{Candidate hot Jupiters for magnetically-driven wind variations} \label{app:table}

\startlongtable
\begin{deluxetable}{llrrr}
\tablecaption{HJs in which  $B_{\di{dip},\di{crit}} < 28 \gauss$, at $P=10 \mbar$ with $\Delta T/T_\di{eq} = 0.1$. Alongside $T_\di{eq}$, estimates of $B_{\di{dip},\di{crit}}$ and $B_{\di{dip},\di{crit}}$  are provided for these choices. %If HJs in this table are observed to magnetic wind variations, $B_{\di{dip},\di{crit},0.1}$ estimates the lower bound of $B_\di{dip}$ and $B_{\phi,\di{crit},0.1}$ estimates  the magnitude of $B_{\phi}$.  
\label{tab:reversal:HJs}
} 
\tablecolumns{5}
\tablehead{
%\colhead{} & 
\colhead{Rank} & 
\colhead{Candidate} & 
\colhead{ $T_\di{eq} / \di{K}$} & 
\colhead{ $B_{\phi,\di{crit},0.1} / \di{G}$} & 
\colhead{ $B_{\di{dip},\di{crit},0.1} / \di{G}$} }
\startdata
1  &              WASP-189 b &     2618 &        129 &          0.9 \\
2$^\dagger$  &               $\dagger \quad$ WASP-12 b &     2578 &        156 &          1 \\
3  &              WASP-178 b &     2366 &        130 &          1 \\
4$^\dagger$  &                $\dagger \quad$ WASP-33 b &     2681 &        149 &          2 \\
5  &              WASP-121 b &     2358 &        153 &          2 \\
6  &             MASCARA-1 b &     2545 &        134 &          3 \\
7  &               WASP-78 b &     2194 &        139 &          3 \\
8  &              HAT-P-70 b &     2551 &        133 &          3 \\
9  &            HD 85628 A b &     2403 &        128 &          3 \\
10 &               HATS-68 b &     1743 &        177 &          3 \\
11 &               WASP-76 b &     2182 &        145 &          3 \\
12 &               WASP-82 b &     2188 &        132 &          4 \\
13 &           HD 202772 A b &     2132 &        125 &          4 \\
14 &             Kepler-91 b &     2037 &        105 &          4 \\
15 &  TOI-1431 b/MASCARA-5 b &     2370 &        129 &          4 \\
16 &              HAT-P-65 b &     1953 &        138 &          5 \\
17 &              WASP-100 b &     2201 &        131 &          6 \\
18 &              WASP-187 b &     1952 &        116 &          6 \\
19 &               HATS-67 b &     2195 &        146 &          6 \\
20 &             WASP-87 A b &     2311 &        139 &          6 \\
21 &               HATS-56 b &     1902 &        122 &          7 \\
22 &               HATS-40 b &     2121 &        126 &          7 \\
23 &               KELT-18 b &     2082 &        130 &          7 \\
24 &              HAT-P-57 b &     2198 &        130 &          7 \\
25 &               HATS-26 b &     1925 &        130 &          7 \\
26$^\dagger$ &                $\dagger \quad$ HAT-P-7 b &     2192 &        134 &          7 \\
27 &               WASP-48 b &     2058 &        139 &          7 \\
28 &                KOI-13 b &     2550 &        139 &          8 \\
29 &              HAT-P-49 b &     2127 &        128 &          9 \\
30 &              WASP-142 b &     1992 &        139 &         11 \\
31 &              WASP-111 b &     2121 &        133 &         11 \\
32 &               WASP-90 b &     1840 &        124 &         12 \\
33 &              HAT-P-66 b &     1900 &        130 &         12 \\
34 &              Qatar-10 b &     1955 &        145 &         13 \\
35 &               KELT-11 b &     1711 &        113 &         13 \\
36 &              HAT-P-33 b &     1839 &        130 &         14 \\
37 &               HATS-35 b &     2033 &        140 &         14 \\
38 &              HAT-P-60 b &     1786 &        119 &         15 \\
39 &               Qatar-7 b &     2052 &        141 &         15 \\
40 &               CoRoT-1 b &     2007 &        146 &         15 \\
41$^\dagger$ &              $\dagger \quad$ Kepler-76 b &     2145 &        142 &         15 \\
42 &                K2-260 b &     1985 &        132 &         15 \\
43 &               WASP-71 b &     2064 &        128 &         15 \\
44 &               WASP-88 b &     1763 &        119 &         16 \\
45 &              WASP-172 b &     1745 &        114 &         16 \\
46 &              WASP-159 b &     1811 &        120 &         17 \\
47 &            Kepler-435 b &     1731 &        109 &         18 \\
48 &               HATS-31 b &     1837 &        128 &         19 \\
49 &              WASP-122 b &     1962 &        147 &         19 \\
50 &              HAT-P-32 b &     1841 &        142 &         19 \\
51 &              HAT-P-23 b &     2133 &        148 &         20 \\
52 &               WASP-92 b &     1879 &        137 &         20 \\
53 &               HATS-64 b &     1800 &        119 &         21 \\
54 &               WASP-19 b &     2060 &        160 &         21 \\
55 &              KELT-4 A b &     1827 &        133 &         21 \\
56 &              CoRoT-21 b &     2041 &        126 &         22 \\
57 &                HATS-9 b &     1913 &        135 &         23 \\
58 &              HAT-P-69 b &     1980 &        118 &         23 \\
59 &           OGLE-TR-132 b &     1981 &        138 &         24 \\
60 &               HATS-24 b &     2091 &        148 &         25 \\
61 &           Kepler-1658 b &     2185 &        110 &         25 \\
62 &               TOI-954 b &     1704 &        109 &         26 \\
63 &              WASP-114 b &     2028 &        142 &         26 \\
64 &               TOI-640 b &     1749 &        120 &         27 \\
65 &              WASP-153 b &     1712 &        128 &         27 \\
 \enddata
\tablenotetext{\dagger}{More accurate estimates in \cref{tab:Bphi:Bdip:estimates}.}
%\tablecomments{Email corresponding author for full list.}
\end{deluxetable}
~ % <- need otherwise table crops

\end{document}